\documentclass[aps,prb,twocolumn,superscriptaddress,showpacs,floatfix]{revtex4}
\usepackage{graphicx}
\usepackage{amsmath}
\usepackage{amssymb}
\usepackage{url}

\graphicspath{{.}{../}{./eps/}}

\begin{document}
\title{Unusual localisation effects in quantum percolation}
\author{Gerald Schubert}
\affiliation{Institut f\"ur Physik, Ernst-Moritz-Arndt Universit\"at
  Greifswald, 17487 Greifswald, Germany}
\author{Alexander Wei{\ss}e}
\affiliation{School of Physics, The University of New South Wales,
  Sydney, NSW 2052, Australia}
\author{Holger Fehske}
\affiliation{Institut f\"ur Physik, Ernst-Moritz-Arndt Universit\"at
  Greifswald, 17487 Greifswald, Germany}
\date{\today}

\begin{abstract}
  We present a detailed study of the quantum site percolation problem
  on simple cubic lattices, thereby focussing on the statistics of the
  local density of states and the spatial structure of the single
  particle wavefunctions. Using the Kernel Polynomial Method we refine
  previous studies of the metal-insulator transition and demonstrate
  the non-monotonic energy dependence of the quantum percolation
  threshold. Remarkably, the data indicates a ``fragmentation'' of the
  spectrum into extended and localised states. In addition, the
  observation of a chequerboard-like structure of the wavefunctions at
  the band centre can be interpreted as anomalous localisation.
\end{abstract}


\maketitle

Disordered structures attracted continuing 
interest over the last decades, and besides the Anderson localisation 
problem\cite{An58} quantum
percolation\cite{KE72,MDS95} is one of the classical subjects of this
field. Current applications concern e.g. transport properties of doped
semiconductors\cite{ITA94} and granular metals\cite{FIS04}, 
metal-insulator transition in two-dimensional n-GaAs 
heterostructures\cite{Saea04}, 
wave propagation through binary inhomogeneous media\cite{AL92},
superconductor-insulator and (integer) quantum Hall
transitions\cite{DMA04,SMK03}, or the dynamics of atomic Fermi-Bose
mixtures\cite{SKSZL04}. Another important example is the
metal-insulator transition in perovskite manganite films and the related
colossal magnetoresistance effect, which in the meantime are believed
to be inherently percolative.\cite{BSLMDSS02}
   
In disordered solids the percolation problem is characterised by the
interplay of pure classical and quantum effects. Besides the question
of finding a percolating path of ``accessible'' sites through a given
lattice the quantum nature of the electrons imposes further
restrictions on the existence of extended states and, consequently, of
a finite DC-conductivity.  As a particularly simple model describing
this situation we consider a tight-binding one-electron Hamiltonian, 
\begin{equation}\label{H_bm}
  {H} =  \sum_{i=1}^{N} \epsilon_i {c}_i^{\dag} {c}_i^{} 
  - t \sum_{\langle ij \rangle}({c}_i^{\dag} {c}_j^{} + \text{H.c.})\,,
\end{equation}
on a simple cubic lattice with $N=L^3$ sites and random on-site energies 
$\epsilon_i$ drawn from the bimodal distribution
\begin{equation}
  p(\epsilon_i) = p\,\delta(\epsilon_i-\epsilon_A) +
  (1-p)\, \delta(\epsilon_i-\epsilon_B)\,.
\end{equation}
The two energies $\epsilon_A$ and $\epsilon_B$ could, for instance,
represent the potential landscape of a binary alloy A$_p$B$_{1-p}$,
where each site is occupied by an A or B atom with probability $p$ or
$1-p$, respectively.  In the limit $\Delta=(\epsilon_B-\epsilon_A)
\rightarrow \infty$ the wavefunction of the $A$ sub-band vanishes
identically on the $B$-sites, making them completely inaccessible for
the quantum particles.  We then arrive at a situation where
non-interacting electrons move on a random ensemble of lattice points,
which, depending on $p$, may span the entire lattice or not. The
corresponding Hamiltonian reads
\begin{equation}\label{H_sc}
  {H} = - t \!\sum_{\langle ij \rangle \in A}\!
  ({c}_i^{\dag} {c}_j + \text{H.c.})\,,
\end{equation}  
where the summation extends over nearest-neighbour $A$-sites only and,
without loss of generality, $\epsilon_A$ is chosen to be zero.

Within the classical percolation scenario the percolation threshold $p_c$
is defined by the occurrence of an infinite cluster $A_\infty$ of
adjacent $A$ sites. For the simple cubic lattice this site-percolation
threshold is $p_c=0.3117$.\cite{HS81} In the quantum case, the
multiple scattering of the particles at the irregular boundaries of
the cluster can suppress the wavefunction, in particular, within
narrow channels or close to dead ends of the cluster.  Hence, this
type of disorder can lead to absence of diffusion due to localisation,
even if there is a classical percolating path through the crystal.  On
the other hand, for finite $\Delta$ the tunnelling between $A$ and $B$
sites may cause a finite DC-conductivity although the $A$ sites are not
percolating.  Naturally, the question arises whether the quantum
percolation threshold $p_q$, given by the probability above which an
extended wavefunction exists within the $A$ sub-band, is larger or
smaller than $p_c$.  Previous results~\cite{SLG92} for finite values
of $\Delta$ indicate that the tunnelling effect has a marginal
influence on the percolation threshold as soon as $\Delta\gg 4tD$,
where $D$ denotes the spatial dimension of the hypercubic lattice.

In the theoretical investigation of disordered systems it turned out
that distribution functions for the random quantities take the centre
stage.\cite{An58,AAT73} The distribution $f(\rho_i(E))$ of the local
density of states (LDOS),
\begin{equation} \label{LDOS}
  \rho_i(E) = \sum\limits_{n=1}^{N}
  | \psi_n ({\bf r}_i)|^2\, \delta(E-E_n)\,,
\end{equation}
is particularly suited because $\rho_i(E)$ measures the local
amplitude of the wavefunction at site ${\bf r}_i$. It therefore
contains direct information about the localisation properties.  In
contrast to the (arithmetically averaged) mean DOS, $\rho_{\rm me}(E)
= \langle\rho_i(E)\rangle$, the LDOS becomes critical at the
localisation transition.\cite{HT94,DPN03} The probability density
$f(\rho_i(E))$ was found to have essentially different properties for
extended and localised phases.\cite{MF94} For an extended state at
energy $E$ the amplitude of the wavefunctions is more or less uniform.
Accordingly $f(\rho_i(E))$ is sharply peaked and symmetric about
$\rho_{\rm me}(E)$.  On the other hand, if states become localised,
the wavefunction has considerable weight only on a few sites.  In this
case the LDOS strongly fluctuates throughout the lattice and the
corresponding LDOS distribution is very asymmetric and 
has a long tail.
Above the localisation transition the distribution of the
LDOS is singular, i.e., mainly concentrated at $\rho_i=0$.
Nevertheless the rare but large LDOS-values dominate the mean DOS
$\rho_{\rm me}(E)$, which therefore cannot be taken as a good
approximation of the most probable value of the LDOS. Such systems are
referred to as ``non-self-averaging''. Of course, for practical
calculations the recording of entire distributions is a bit
inconvenient. Instead the mean DOS $\rho_{\rm me}(E)$ together
with the (geometrically averaged) so-called ``typical'' DOS,
$\rho_{\rm ty}(E) = \exp\langle \ln\rho_i(E)\rangle$, is frequently
used to monitor the transition from extended to localised states. The
typical DOS puts sufficient weight on small values of $\rho_i$ and a
comparison to $\rho_{\rm me}(E)$ therefore allows to detect the
localisation transition.  This has been shown for the pure Anderson
model\cite{DPN03,SWF03,ASWBF04} and for even more complex situations,
where the effects of correlated disorder\cite{SWF04b},
electron-electron interaction\cite{DK97,BHV04} or electron-phonon
coupling\cite{BF02,BAF04} were taken into account.

In this paper we employ the typical-DOS concept to analyse the nature
of the eigenstates (extended or localised) of the Hamiltonians
\eqref{H_bm} and~\eqref{H_sc}. Using the Kernel Polynomial
Method\cite{SRVK96,SWF03}, an efficient high-resolution Chebyshev
expansion technique, in a first step, we calculate the LDOS for a
large number of samples, $K_r$, and sites, $K_s$. The mean DOS is then
simply given by
\begin{equation}
  \rho_{\text{me}}(E)  = \frac{1}{K_r K_s} \smash{
   \sum\limits_{k=1}^{K_r}\sum\limits_{i=1}^{K_s}
   \rho_i(E)} \,, \label{rhoav}
\end{equation}
whereas the typical DOS is obtained from the geometric average
\begin{equation}
  \rho_{\text{ty}}(E)  = \exp \left(\frac{1}{K_r K_s} \smash{
   \sum\limits_{k=1}^{K_r}\sum\limits_{i=1}^{K_s}
   \ln\bigl(\rho_i(E)\bigr)} \right) \,. \label{rhoty}
\end{equation}
We classify a state at energy $E$ with $\rho_{\rm me}(E)\neq 0$ as
localised if $\rho_{\rm ty}(E)= 0$ and as extended if $\rho_{\rm
  ty}(E)\neq 0$.

\begin{figure}
  \centering 
  \includegraphics[width=0.95\linewidth]{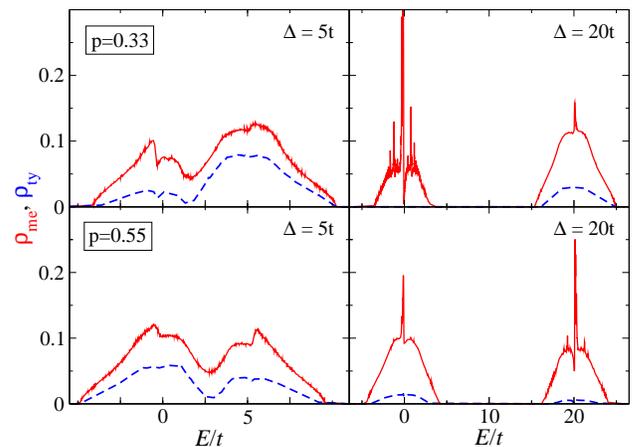}
   \caption{(Colour online) Mean (solid line) and typical (dashed line) 
     DOS of the Hamiltonian~\eqref{H_bm} on a $50^3$ lattice with
     periodic boundary conditions (PBC). Results are obtained using
     $M=32768$ Chebyshev moments and $K_s\times K_r=32\times 32$
     realisations.}\label{endl_delta}
\end{figure}

\begin{figure}
  \centering 
  \includegraphics[width=0.95\linewidth]{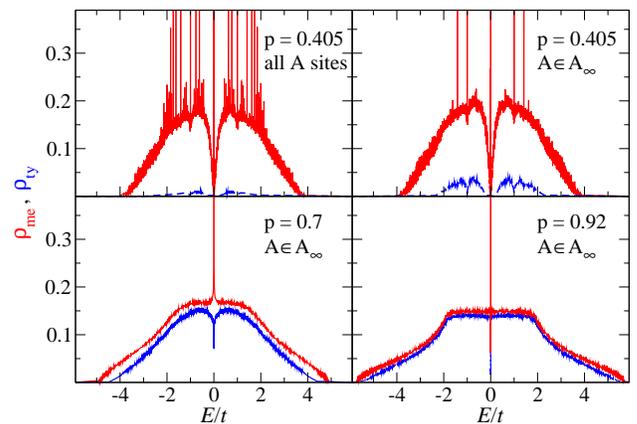}
  \caption{(Colour online) Mean (solid line) and typical (dashed line)
    DOS for the quantum percolation model in the limit
    $\Delta\to\infty$.  While in the upper left panel all $A$-sites
    are taken into account, the other three panels show data for the
    restricted model~\eqref{H_sc} on the spanning cluster $A_\infty$
    only (note that  $\rho_{\rm ty}$ is smaller in the former case
    because there are more sites with vanishing amplitude 
    of the wavefunction). System sizes were 
    adapted to ensure that $A_\infty$ always
    contains the same number of sites, i.e., $N=57^3$ for $p=0.405$,
    $46^3$ for $p=0.70$, and $42^3$ for $p=0.92$.  Again we used
    $M=32768$ and $K_s\times K_r=32\times 32$.  }\label{Perc}
\end{figure}

Before discussing possible localisation phenomena let us investigate
the behaviour of the mean DOS for the quantum percolation
models~\eqref{H_bm} and~\eqref{H_sc}.  Figure~\ref{endl_delta} shows
that as long as $\epsilon_A$ and $\epsilon_B$ do not differ too much
there exists an asymmetric (if $p\neq 0.5$) but still connected
electronic band.\cite{SLG92} At about $\Delta \simeq 4tD$ this band
separates into two sub-bands centred at $\epsilon_A$ and $\epsilon_B$,
respectively.  The most prominent feature in the split-band regime is
the series of spikes at discrete energies within the band.  As an
obvious guess, we might attribute these spikes to eigenstates on
islands of $A$ or $B$ sites being isolated from the main
cluster.\cite{KE72,BA96} 
It turns out, however, that some of the
spikes persist, even if we neglect all finite clusters and restrict
the calculation to the spanning cluster of $A$ sites, $A_\infty$.
This is illustrated in the upper panels of Fig.~\ref{Perc}, where we
compare the DOS of the models \eqref{H_bm} [at $\Delta\to\infty$]
and~\eqref{H_sc}.  Increasing the concentration of accessible sites
the mean DOS of the spanning cluster is evocative of the DOS of the
simple cubic lattice, but even at large values of $p$ a sharp peak
structure remains at $E=0$ (cf. Fig.~\ref{Perc}, lower panels).


\begin{figure}
  \centering 
  \includegraphics[width=0.95\linewidth]{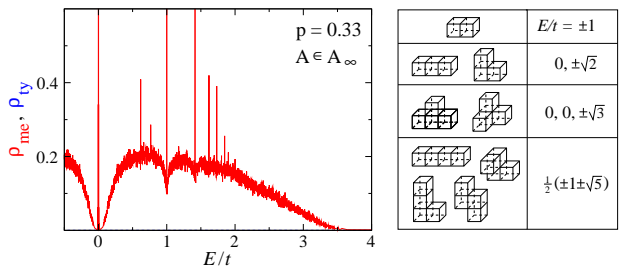}
  \caption{(Colour online) Left: Mean (solid line) and typical (dashed line)
    DOS for the model~\eqref{H_sc} with $p=0.33$ on a $100^3$ lattice
    (PBC, $M=32768$). Data obtained from an average over $100$ random
    initialisations of sites for $K_r=100$ realisations of disorder.
    Note that $\rho_{ty} <10^{-5}$ holds in the whole band.  Right:
    Some cluster configurations related to the special energies at
    which the peaks in $\rho_{\text{me}}$ occur.}\label{Cl_Tab}
\end{figure}

To elucidate this effect, which partially is not accounted 
for in the literature\cite{KE72,OOM80,SEG87,SLG92}, 
in more detail, in Fig.~\ref{Cl_Tab} we fixed $p$ at $0.33$, shortly
above the classical percolation threshold, and increased the
ensemble size. Besides the most dominant peaks at $E/t = 0, \pm 1,
\pm\sqrt{2}$, in addition we can resolve distinct spikes at $E/t =
\frac{1}{2}\left(\pm 1\pm \sqrt{5}\right), \pm \sqrt{3}, \pm\sqrt{ 2\pm 
\sqrt{2}}, \ldots$ These special energies coincide with the
eigenvalues of the tight-binding model on small clusters of the
geometries shown in the right part of Fig.~\ref{Cl_Tab}. In accordance
with Refs.~\onlinecite{KE72} and~\onlinecite{CCFST86} we can thus
argue that the wavefunctions, which correspond to these special
energies, are localised on some ``dead ends'' of the spanning cluster.

The assumption that the distinct peaks correspond to localised
wavefunctions is corroborated by the fact that the typical DOS
vanishes or, at least, shows a dip at these energies.  Occurring also
for finite $\Delta$ (Fig.~\ref{endl_delta}), this effect becomes more
pronounced as $\Delta\to\infty$ and in the vicinity of the classical
percolation threshold $p_c$.  From the study of the Anderson
model~\cite{An58} we know that localisation leads at first to a
narrowing of the energy window containing extended states.  The
corresponding mobility edges have been mapped out with high
precision.\cite{MK83,ASWBF04} For the percolation problem, in
contrast, with decreasing $p$ the typical DOS indicates both
localisation from the band edges and localisation at particular
energies within the band.  Since finite cluster wavefunctions like
those shown in Fig.~\ref{Cl_Tab} can be constructed for numerous
other, less probable geometries~\cite{Sc03}, \textcite{CCFST86} argued
that an infinite discrete series of such spikes might exist within the
spectrum. The picture of localisation in the quantum percolation model
is then quite remarkable. If we generalise our numerical data for the
peaks at $E=0$ and $E/t=\pm 1$, it seems as if there is an infinite
discrete set of energies with localised wavefunctions, which is dense
within the entire spectrum. In between there are continua of
delocalised states, but to avoid mixing, their density goes to zero
close to the localised states. Facilitated by the large special weight
of the peak (up to 10\% close to $p_c$) this is clearly observed at
$E=0$, and we suspect similar behaviour at $E/t=\pm 1$. For the other
discrete spikes the resolution of our numerical data is still too poor
and the system size might be even too small to draw a definite conclusion.

\begin{figure}
  \centering 
  \includegraphics[width=0.95\linewidth]{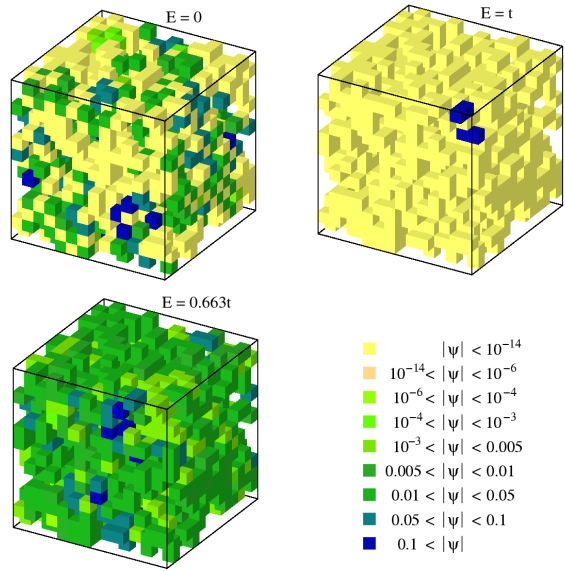}
  \caption{(Colour online) Amplitudes of the wavefunction, 
    $|\psi_n({\bf r}_i)|$, of the quantum percolation
    model~\eqref{H_sc} on $A_\infty$ of a $14^3$ lattice with
    occupation probability $p=0.45$.  Exact diagonalisation results
    are presented at three characteristic energies $E/t=0$, 1, and
    0.66 corresponding to anomalously localised, localised, and
    delocalised states, respectively.  }\label{CharZst1}
\end{figure}

In order to understand the internal structure of the extended and
localised states we calculated the amplitudes of the wavefunction at
specific energies for a random sample of the quantum percolation
model~\eqref{H_sc} restricted to $A_\infty$.  
Figure~\ref{CharZst1} visualises the spatial
variation of $|\psi_n({\bf r_i})|$ on a $14^3$ lattice with PBC and an
occupation probability $p=0.45$ well above the classical percolation
threshold. The figure clearly indicates that the state with $E/t=0.66$
is extended, i.e. the spanning cluster is quantum mechanically
``transparent''. On the contrary, at $E=t$, the wavefunction is
completely localised on a finite region of the spanning cluster.  Here
the scattering of the particle at the random surface of the spanning
cluster results in states, where the wavefunction vanishes identically
except for some finite domains on loose ends 
(like those shown in Fig.~\ref{Cl_Tab}), where it takes the values ($\pm
1, \mp 1$), ($\pm 1,\mp 1, 0,\mp 1,\pm 1$), $\ldots$  Note that these
regions are part of the spanning cluster, connected to the other sites
by a site with wavefunction amplitude zero. A particularly interesting
behaviour is observed at $E=0$.  The eigenstate $E=0$ is highly
degenerate and we can form wavefunctions that span the entire lattice
in a chequerboard structure with zero and non-zero amplitudes (see
Fig.~\ref{CharZst1}).  Although these states are extended in the sense
that they are not confined to some region of the cluster, they are
localised in the sense that they do not contribute to the
DC-conductivity. This is caused by the alternating structure which
suppresses the nearest-neighbour hopping, and in spite of the high
degeneracy, the current matrix element between different $E=0$ states
is zero.  Hence, having properties of both classes of states these
states are called anomalously localised.\cite{SAH82,ITA94} The
chequerboard structure is also observed for the hypercubic lattice in
2D but with reduced spectral weight, compared to the 3D case.  Another
indication for the robustness of this feature is its persistence for
mismatching boundary conditions, e.g., periodic (antiperiodic)
boundary conditions for odd (even) values of the linear extension $L$. 
In these cases the chequerboard is matched to itself by a
manifold of sites with vanishing amplitude.

\begin{figure}
  \centering 
  \includegraphics[width=0.95\linewidth]{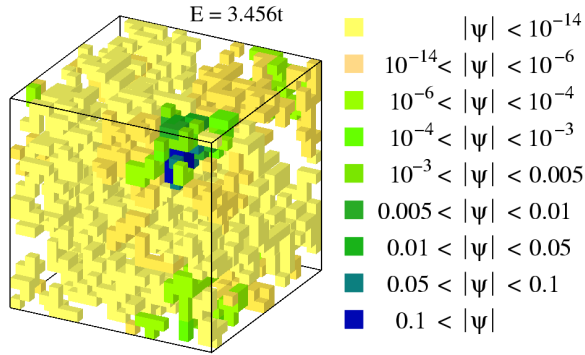}
  \caption{(Colour online) Amplitudes $|\psi_n({\bf r}_i)|$ of a localised  
    state on a $21^3$ lattice with $p=0.33$ obtained by exact
    diagonalisation of the model~\eqref{H_sc}.  $E/t=3.456$ was chosen
    in order to avoid any of the special cluster configurations
    discussed above.}\label{CharZst2}
\end{figure}

In the past most of the methods used in numerical studies of Anderson
localisation have also been applied to the percolation
models~\eqref{H_bm} or~\eqref{H_sc} in order to determine the quantum
percolation threshold $p_q$, defined as the probability $p$ below
which all states are localised (see, e.g.,
Refs.~\onlinecite{MDS95,KN90} and references therein). However, so far
the results for $p_q$ are far less precise than, e.g., the values of
the critical disorder reported for the Anderson model. For the simple
cubic lattice numerical estimates of quantum site-percolation
thresholds range from 0.4 to 0.5. In
Figs.~\ref{endl_delta}-\ref{Cl_Tab} we presented data for $\rho_{\rm
  ty}$ which shows that $p_q>p_c$. In fact, within numerical accuracy,
we found $\rho_{\rm ty}=0$ for $p=0.33>p_c$.  Figure~\ref{CharZst2}
displays the amplitude of a typical state obtained by exact
diagonalisation for a fixed realisation of disorder.  Bearing in mind
that we have used PBC, the support of the wavefunction appears to be a
finite (connected) region of the spanning cluster $A_\infty$, i.e.,
the state is clearly localised.

\begin{figure}[t!]
  \centering 
  \includegraphics[width=\linewidth]{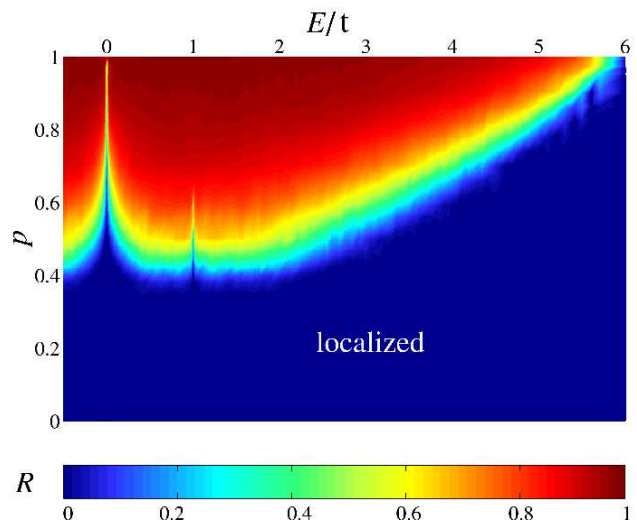}
  \caption{(Colour online) Normalised typical DOS $
R=\rho_{\rm ty}/\rho_{\rm me}$ 
    in the concentration-energy plane for the model~\eqref{H_sc} on a
    $50^3$ lattice for $p\ge 0.5$ and a $100^3$ lattice for $p<0.5$.
    We used $M=16384$ Chebyshev moments and averaged over $K_s\times
    K_r=32\times 32$ realisations.
    \label{Soukoulis}
    }
\end{figure}

To get a more detailed picture we calculated the normalised typical
DOS, $R(p,E)=\rho_{\rm ty}/\rho_{\rm me}$, in the whole
concentration-energy plane.  Figure~\ref{Soukoulis} presents such kind
of phase diagram of the quantum percolation model~\eqref{H_sc}. The
data supports a finite quantum percolation threshold $p_q\gtrsim
0.4>p_c$ (cf. also Refs.~\onlinecite{KN90,SLG92,KSHS97,KO99}), but as
the discussion above indicated, for $E=0$ and $E=\pm t$ the critical
value $p_q(E)$ is $1$, and the same may hold for the set of other
``special'' energies.  The transition line between localised and
delocalised states, $p_q(E)$, might thus be a rather irregular
(fractal?) function.  On the basis of our numerical treatment,
however, we are not in the position to answer this question with full
rigour. 
\begin{figure}[ht!]
  \centering 
  \includegraphics[width=0.95\linewidth]{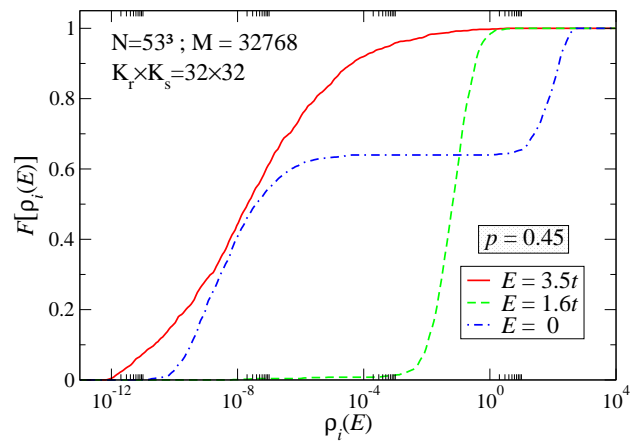}
  \caption{(Colour online) Characteristic probability distributions 
    of the LDOS.}\label{DistFkt}
\end{figure}

Finally let us come back to the characterisation of extended and
localised states in terms of distribution functions.
Figure~\ref{DistFkt} displays the probability distribution,
\begin{equation}
F[\rho_i(E)]=\int_0^{\rho_i(E)} f(\rho^\prime_i(E))\,d\rho^\prime_i(E)\,,
\end{equation}
for three typical energies $E/t=3.5$, 1.6, and 0 corresponding to
localised, extended, and anomalously localised states at $p=0.45$,
respectively. The differences in $F[\rho_i(E)]$ are significant. The
slow increase of $F[\rho_i(E)]$ observed for localised states
corresponds to an extremely broad LDOS-distribution, with a very small
most probable (or typical) value of $\rho_i(E)$.  This is in agreement
with the findings for the Anderson model.~\cite{SWF03,ASWBF04}
Accordingly the jump-like increase found for extended states is
related to an extremely narrow distribution of the LDOS centred around
the mean DOS, where $\rho_{\rm me}$ coincides with the most probable
value.  At $E=0$, the probability distribution exhibits two steps,
leading to a bimodal distribution density. Here the first (second)
maximum is related to sites with a small (large) amplitude of the
wavefunction -- a feature that substantiates the chequerboard 
structure discussed above.

To summarise, we have demonstrated the value and power of the
probability distribution approach to quantum percolation. As for
standard Anderson localisation the typical density of states can serve
as a kind of order parameter differentiating between extended and
localised states.  Our numerical data corroborates previous results in
favour of a quantum percolation threshold $p_q>p_c$ and a
fragmentation of the spectrum into extended and localised states.
The latter refers to a discrete but dense set of localised states
separated by continua of delocalised states. Accordingly the function
$p_q(E)$ is rather irregular.  At the band centre, so-called anomalous
localisation is observed, which manifests itself in a
chequerboard-like structure of the wavefunction.  Even though the
Kernel Polynomial Method allows for the study of very large clusters
with high energy resolution, the quantum percolation problem certainly
deserves further investigation.

It is a pleasure to acknowledge useful discussions with A.~Alvermann,
F.X.~Bronold, G.~Wellein and W.~Weller.  Special thanks go to LRZ
M\"unchen, NIC J\"ulich and HLRN (Zuse-Institut Berlin) for granting
resources on their supercomputing facilities.


\begin{thebibliography}{35}
\expandafter\ifx\csname natexlab\endcsname\relax\def\natexlab#1{#1}\fi
\expandafter\ifx\csname bibnamefont\endcsname\relax
  \def\bibnamefont#1{#1}\fi
\expandafter\ifx\csname bibfnamefont\endcsname\relax
  \def\bibfnamefont#1{#1}\fi
\expandafter\ifx\csname citenamefont\endcsname\relax
  \def\citenamefont#1{#1}\fi
\expandafter\ifx\csname url\endcsname\relax
  \def\url#1{\texttt{#1}}\fi
\expandafter\ifx\csname urlprefix\endcsname\relax\def\urlprefix{URL }\fi
\providecommand{\bibinfo}[2]{#2}
\providecommand{\eprint}[2][]{\url{#2}}

\bibitem[{\citenamefont{Anderson}(1958)}]{An58}
\bibinfo{author}{\bibfnamefont{P.~W.} \bibnamefont{Anderson}},
  \bibinfo{journal}{Phys. Rev.} \textbf{\bibinfo{volume}{109}},
  \bibinfo{pages}{1492} (\bibinfo{year}{1958}).

\bibitem[{\citenamefont{Kirkpatrick and Eggarter}(1972)}]{KE72}
\bibinfo{author}{\bibfnamefont{S.}~\bibnamefont{Kirkpatrick}} \bibnamefont{and}
  \bibinfo{author}{\bibfnamefont{T.~P.} \bibnamefont{Eggarter}},
  \bibinfo{journal}{Phys. Rev. B} \textbf{\bibinfo{volume}{6}},
  \bibinfo{pages}{3598} (\bibinfo{year}{1972}).

\bibitem[{\citenamefont{Mookerjee et~al.}(1995)\citenamefont{Mookerjee,
  Dasgupta, and Saha}}]{MDS95}
\bibinfo{author}{\bibfnamefont{A.}~\bibnamefont{Mookerjee}},
  \bibinfo{author}{\bibfnamefont{I.}~\bibnamefont{Dasgupta}}, \bibnamefont{and}
  \bibinfo{author}{\bibfnamefont{T.}~\bibnamefont{Saha}},
  \bibinfo{journal}{Int. J. Mod. Phys. B} \textbf{\bibinfo{volume}{9}},
  \bibinfo{pages}{2989} (\bibinfo{year}{1995}).

\bibitem[{\citenamefont{Inui et~al.}(1994)\citenamefont{Inui, Trugman, and
  Abrahans}}]{ITA94}
\bibinfo{author}{\bibfnamefont{M.}~\bibnamefont{Inui}},
  \bibinfo{author}{\bibfnamefont{S.~A.} \bibnamefont{Trugman}},
  \bibnamefont{and} \bibinfo{author}{\bibfnamefont{E.}~\bibnamefont{Abrahans}},
  \bibinfo{journal}{Phys. Rev. B} \textbf{\bibinfo{volume}{49}},
  \bibinfo{pages}{3190} (\bibinfo{year}{1994}).




\bibitem[{\citenamefont{Feigel'mann et~al.}(2004)\citenamefont{Feigel'mann,
  Ioselevich, and Skvortsov}}]{FIS04}
\bibinfo{author}{\bibfnamefont{M.~V.} \bibnamefont{Feigel'mann}},
  \bibinfo{author}{\bibfnamefont{A.}~\bibnamefont{Ioselevich}},
  \bibnamefont{and} \bibinfo{author}{\bibfnamefont{M.}~\bibnamefont{Skvortsov}}
  (\bibinfo{year}{2004}),
  \urlprefix\url{http://arXiv.org/abs/cond-mat/0404350}.

\bibitem[{\citenamefont{Das Sarma et~al.}(2004)
 \citenamefont{Das Sarma et al.}}]{Saea04}
\bibinfo{author}{\bibfnamefont{S.} \bibnamefont{Das Sarma}},
  \bibinfo{author}{\bibfnamefont{M. P.}~\bibnamefont{Lilly}},
  \bibinfo{author}{\bibfnamefont{E. H.}~\bibnamefont{Hwang}},
  \bibinfo{author}{\bibfnamefont{L. N.}~\bibnamefont{Pfeiffer}},
   \bibinfo{author}{\bibfnamefont{K. W.}~\bibnamefont{West}},
  \bibnamefont{and} \bibinfo{author}{\bibfnamefont{J. L.}~\bibnamefont{Reno}}
  (\bibinfo{year}{2004}),
  \urlprefix\url{http://arXiv.org/abs/cond-mat/0406655}.

\bibitem[{\citenamefont{Avishai and Luch}(1992)}]{AL92}
\bibinfo{author}{\bibfnamefont{Y.}~\bibnamefont{Avishai}} \bibnamefont{and}
  \bibinfo{author}{\bibfnamefont{J.}~\bibnamefont{Luch}},
  \bibinfo{journal}{Phys. Rev. B} \textbf{\bibinfo{volume}{45}},
  \bibinfo{pages}{1974} (\bibinfo{year}{1992}).

\bibitem[{\citenamefont{Dubi et~al.}(2004)\citenamefont{Dubi, Meir, and
  Avishai}}]{DMA04}
\bibinfo{author}{\bibfnamefont{Y.}~\bibnamefont{Dubi}},
  \bibinfo{author}{\bibfnamefont{Y.}~\bibnamefont{Meir}}, \bibnamefont{and}
  \bibinfo{author}{\bibfnamefont{Y.}~\bibnamefont{Avishai}}
  (\bibinfo{year}{2004}),
  \urlprefix\url{http://arXiv.org/abs/cond-mat/0406008}.

\bibitem[{\citenamefont{Sandler et~al.}(2003)\citenamefont{Sandler, Maei, and
  Kondev}}]{SMK03}
\bibinfo{author}{\bibfnamefont{N.}~\bibnamefont{Sandler}},
  \bibinfo{author}{\bibfnamefont{H.}~\bibnamefont{Maei}}, \bibnamefont{and}
  \bibinfo{author}{\bibfnamefont{J.}~\bibnamefont{Kondev}}
  (\bibinfo{year}{2003}),
  \urlprefix\url{http://arXiv.org/abs/cond-mat/0311484}.

\bibitem[{\citenamefont{Sanpera et~al.}(2004)\citenamefont{Sanpera, Kantian,
  Sanchez-Palencia, Zakrzewski, and Lewenstein}}]{SKSZL04}
\bibinfo{author}{\bibfnamefont{A.}~\bibnamefont{Sanpera}},
  \bibinfo{author}{\bibfnamefont{A.}~\bibnamefont{Kantian}},
  \bibinfo{author}{\bibfnamefont{L.}~\bibnamefont{Sanchez-Palencia}},
  \bibinfo{author}{\bibfnamefont{J.}~\bibnamefont{Zakrzewski}},
  \bibnamefont{and}
  \bibinfo{author}{\bibfnamefont{M.}~\bibnamefont{Lewenstein}}
  (\bibinfo{year}{2004}),
  \urlprefix\url{http://arXiv.org/abs/cond-mat/0402375}.

\bibitem[{\citenamefont{Becker et~al.}(2002)\citenamefont{Becker, Streng,
  Y.~Luo, Damaschke, Shannon, and Samwer}}]{BSLMDSS02}
\bibinfo{author}{\bibfnamefont{T.}~\bibnamefont{Becker}},
  \bibinfo{author}{\bibfnamefont{C.}~\bibnamefont{Streng}},
  \bibinfo{author}{\bibfnamefont{Y.} \bibnamefont{Luo}},
 \bibinfo{author}{\bibfnamefont{V.} \bibnamefont{Moshnyaga}},
  \bibinfo{author}{\bibfnamefont{B.}~\bibnamefont{Damaschke}},
  \bibinfo{author}{\bibfnamefont{N.}~\bibnamefont{Shannon}}, \bibnamefont{and}
  \bibinfo{author}{\bibfnamefont{K.}~\bibnamefont{Samwer}},
  \bibinfo{journal}{Phys. Rev. Lett.} 
  \textbf{\bibinfo{volume}{89}}, \bibinfo{pages}{237203}
  (\bibinfo{year}{2002}).

\bibitem[{\citenamefont{Heermann and Stauffer}(1981)}]{HS81}
\bibinfo{author}{\bibfnamefont{D.~W.} \bibnamefont{Heermann}} \bibnamefont{and}
  \bibinfo{author}{\bibfnamefont{D.}~\bibnamefont{Stauffer}},
  \bibinfo{journal}{Z. Phys. B} \textbf{\bibinfo{volume}{44}},
  \bibinfo{pages}{339} (\bibinfo{year}{1981}).

\bibitem[{\citenamefont{Soukoulis et~al.}(1992)\citenamefont{Soukoulis, Li, and
  Grest}}]{SLG92}
\bibinfo{author}{\bibfnamefont{C.~M.} \bibnamefont{Soukoulis}},
  \bibinfo{author}{\bibfnamefont{Q.}~\bibnamefont{Li}}, \bibnamefont{and}
  \bibinfo{author}{\bibfnamefont{G.~S.} \bibnamefont{Grest}},
  \bibinfo{journal}{Phys. Rev. B} \textbf{\bibinfo{volume}{45}},
  \bibinfo{pages}{7724} (\bibinfo{year}{1992}).

\bibitem[{\citenamefont{Abou-Chacra et~al.}(1973)\citenamefont{Abou-Chacra,
  Anderson, and Thouless}}]{AAT73}
\bibinfo{author}{\bibfnamefont{R.}~\bibnamefont{Abou-Chacra}},
  \bibinfo{author}{\bibfnamefont{P.~W.} \bibnamefont{Anderson}},
  \bibnamefont{and} \bibinfo{author}{\bibfnamefont{D.~J.}
  \bibnamefont{Thouless}}, \bibinfo{journal}{J. Phys. C}
  \textbf{\bibinfo{volume}{6}}, \bibinfo{pages}{1734} (\bibinfo{year}{1973}).

\bibitem[{\citenamefont{Haydock and Te}(1994)}]{HT94}
\bibinfo{author}{\bibfnamefont{R.}~\bibnamefont{Haydock}} \bibnamefont{and}
  \bibinfo{author}{\bibfnamefont{R.~L.} \bibnamefont{Te}},
  \bibinfo{journal}{Phys. Rev. B} \textbf{\bibinfo{volume}{49}},
  \bibinfo{pages}{10845} (\bibinfo{year}{1994}).

\bibitem[{\citenamefont{Dobrosavljevi\'{c}
  et~al.}(2003)\citenamefont{Dobrosavljevi\'{c}, Pastor, and
  Nikoli\'{c}}}]{DPN03}
\bibinfo{author}{\bibfnamefont{V.}~\bibnamefont{Dobrosavljevi\'{c}}},
  \bibinfo{author}{\bibfnamefont{A.~A.} \bibnamefont{Pastor}},
  \bibnamefont{and} \bibinfo{author}{\bibfnamefont{B.~K.}
  \bibnamefont{Nikoli\'{c}}}, \bibinfo{journal}{Europhys. Lett.}
  \textbf{\bibinfo{volume}{62}}, \bibinfo{pages}{76} (\bibinfo{year}{2003}).

\bibitem[{\citenamefont{Mirlin and Fyodorov}(1994)}]{MF94}
\bibinfo{author}{\bibfnamefont{A.~D.} \bibnamefont{Mirlin}} \bibnamefont{and}
  \bibinfo{author}{\bibfnamefont{Y.~V.} \bibnamefont{Fyodorov}},
  \bibinfo{journal}{J. Phys. I (France)} \textbf{\bibinfo{volume}{4}},
  \bibinfo{pages}{655} (\bibinfo{year}{1994}).

\bibitem[{\citenamefont{Schubert et~al.}(2003)\citenamefont{Schubert,
  Wei{\ss}e, and Fehske}}]{SWF03}
\bibinfo{author}{\bibfnamefont{G.}~\bibnamefont{Schubert}},
  \bibinfo{author}{\bibfnamefont{A.}~\bibnamefont{Wei{\ss}e}},
  \bibnamefont{and} \bibinfo{author}{\bibfnamefont{H.}~\bibnamefont{Fehske}}
  (\bibinfo{year}{2003}),
  \urlprefix\url{http://arXiv.org/abs/cond-mat/0309015}.

\bibitem[{\citenamefont{Alvermann et~al.}(2004)\citenamefont{Alvermann,
  Schubert, Wei{\ss}e, Bronold, and Fehske}}]{ASWBF04}
\bibinfo{author}{\bibfnamefont{A.}~\bibnamefont{Alvermann}},
  \bibinfo{author}{\bibfnamefont{G.}~\bibnamefont{Schubert}},
  \bibinfo{author}{\bibfnamefont{A.}~\bibnamefont{Wei{\ss}e}},
  \bibinfo{author}{\bibfnamefont{F.~X.} \bibnamefont{Bronold}},
  \bibnamefont{and} \bibinfo{author}{\bibfnamefont{H.}~\bibnamefont{Fehske}}
  (\bibinfo{year}{2004}),
  \urlprefix\url{http://arXiv.org/abs/cond-mat/0406051}.

\bibitem[{\citenamefont{Schubert et~al.}(2004)\citenamefont{Schubert,
  Wei{\ss}e, and Fehske}}]{SWF04b}
\bibinfo{author}{\bibfnamefont{G.}~\bibnamefont{Schubert}},
  \bibinfo{author}{\bibfnamefont{A.}~\bibnamefont{Wei{\ss}e}},
  \bibnamefont{and} \bibinfo{author}{\bibfnamefont{H.}~\bibnamefont{Fehske}}
  (\bibinfo{year}{2004}),
  \urlprefix\url{http://arXiv.org/abs/cond-mat/0406212}.

\bibitem[{\citenamefont{Dobrosavljevi\'{c} and Kotliar}(1997)}]{DK97}
\bibinfo{author}{\bibfnamefont{V.}~\bibnamefont{Dobrosavljevi\'{c}}}
  \bibnamefont{and} \bibinfo{author}{\bibfnamefont{G.}~\bibnamefont{Kotliar}},
  \bibinfo{journal}{Phys. Rev. Lett.} \textbf{\bibinfo{volume}{78}},
  \bibinfo{pages}{3943} (\bibinfo{year}{1997}).

\bibitem[{\citenamefont{Byczuk et~al.}(2004)\citenamefont{Byczuk, Hofstetter,
  and Vollhardt}}]{BHV04}
\bibinfo{author}{\bibfnamefont{K.}~\bibnamefont{Byczuk}},
  \bibinfo{author}{\bibfnamefont{W.}~\bibnamefont{Hofstetter}},
  \bibnamefont{and} \bibinfo{author}{\bibfnamefont{D.}~\bibnamefont{Vollhardt}}
  (\bibinfo{year}{2004}),
  \urlprefix\url{http://arXiv.org/abs/cond-mat/0403765}.

\bibitem[{\citenamefont{Bronold and Fehske}(2002)}]{BF02}
\bibinfo{author}{\bibfnamefont{F.~X.} \bibnamefont{Bronold}} \bibnamefont{and}
  \bibinfo{author}{\bibfnamefont{H.}~\bibnamefont{Fehske}},
  \bibinfo{journal}{Phys. Rev. B} \textbf{\bibinfo{volume}{66}},
  \bibinfo{pages}{073102} (\bibinfo{year}{2002}).

\bibitem[{\citenamefont{Bronold et~al.}(2004)\citenamefont{Bronold, Alvermann,
  and Fehske}}]{BAF04}
\bibinfo{author}{\bibfnamefont{F.~X.} \bibnamefont{Bronold}},
  \bibinfo{author}{\bibfnamefont{A.}~\bibnamefont{Alvermann}},
  \bibnamefont{and} \bibinfo{author}{\bibfnamefont{H.}~\bibnamefont{Fehske}},
  \bibinfo{journal}{Philos. Mag. B} \textbf{\bibinfo{volume}{1}},
  \bibinfo{pages}{63} (\bibinfo{year}{2004}).

\bibitem[{\citenamefont{Silver et~al.}(1996)\citenamefont{Silver, R\"oder,
  Voter, and Kress}}]{SRVK96}
\bibinfo{author}{\bibfnamefont{R.~N.} \bibnamefont{Silver}},
  \bibinfo{author}{\bibfnamefont{H.}~\bibnamefont{R\"oder}},
  \bibinfo{author}{\bibfnamefont{A.~F.} \bibnamefont{Voter}}, \bibnamefont{and}
  \bibinfo{author}{\bibfnamefont{D.~J.} \bibnamefont{Kress}},
  \bibinfo{journal}{J. of Comp. Phys.} \textbf{\bibinfo{volume}{124}},
  \bibinfo{pages}{115} (\bibinfo{year}{1996}).

\bibitem[{\citenamefont{Berkovits and Avishai}(1996)}]{BA96}
\bibinfo{author}{\bibfnamefont{R.}~\bibnamefont{Berkovits}} \bibnamefont{and}
  \bibinfo{author}{\bibfnamefont{Y.}~\bibnamefont{Avishai}},
  \bibinfo{journal}{Phys. Rev. B} 
  \textbf{\bibinfo{volume}{53}}, \bibinfo{pages}{R16125}
  (\bibinfo{year}{1996}).

\bibitem[{\citenamefont{Odagaki et~al.}(1980)\citenamefont{Odagaki, Ogita, and
  Matsuda}}]{OOM80}
\bibinfo{author}{\bibfnamefont{T.}~\bibnamefont{Odagaki}},
  \bibinfo{author}{\bibfnamefont{N.}~\bibnamefont{Ogita}}, \bibnamefont{and}
  \bibinfo{author}{\bibfnamefont{H.}~\bibnamefont{Matsuda}},
  \bibinfo{journal}{J. Phys. C} \textbf{\bibinfo{volume}{13}},
  \bibinfo{pages}{189} (\bibinfo{year}{1980}).

\bibitem[{\citenamefont{Soukoulis et~al.}(1987)\citenamefont{Soukoulis,
  Economou, and Grest}}]{SEG87}
\bibinfo{author}{\bibfnamefont{C.~M.} \bibnamefont{Soukoulis}},
  \bibinfo{author}{\bibfnamefont{E.~N.} \bibnamefont{Economou}},
  \bibnamefont{and} \bibinfo{author}{\bibfnamefont{G.~S.} \bibnamefont{Grest}},
  \bibinfo{journal}{Phys. Rev. B} \textbf{\bibinfo{volume}{36}},
  \bibinfo{pages}{8649} (\bibinfo{year}{1987}).

\bibitem[{\citenamefont{Chayes et~al.}(1986)\citenamefont{Chayes, Chayes,
  Franz, Sethna, and Trugman}}]{CCFST86}
\bibinfo{author}{\bibfnamefont{J.~T.} \bibnamefont{Chayes}},
  \bibinfo{author}{\bibfnamefont{L.}~\bibnamefont{Chayes}},
  \bibinfo{author}{\bibfnamefont{J.~R.} \bibnamefont{Franz}},
  \bibinfo{author}{\bibfnamefont{J.~P.} \bibnamefont{Sethna}},
  \bibnamefont{and} \bibinfo{author}{\bibfnamefont{S.~A.}
  \bibnamefont{Trugman}}, \bibinfo{journal}{J. Phys. A}
  \textbf{\bibinfo{volume}{19}}, \bibinfo{pages}{L1173} (\bibinfo{year}{1986}).

\bibitem[{\citenamefont{{Mac Kinnon} and Kramer}(1983)}]{MK83}
\bibinfo{author}{\bibfnamefont{A.}~\bibnamefont{{Mac Kinnon}}}
  \bibnamefont{and} \bibinfo{author}{\bibfnamefont{B.}~\bibnamefont{Kramer}},
  \bibinfo{journal}{Z. Phys. B} \textbf{\bibinfo{volume}{53}},
  \bibinfo{pages}{1} (\bibinfo{year}{1983}).

\bibitem[{\citenamefont{Schubert}(2003)}]{Sc03}
\bibinfo{author}{\bibfnamefont{G.}~\bibnamefont{Schubert}},
  \bibinfo{type}{diploma thesis}, \bibinfo{school}{Universit\"{a}t Bayreuth}
  (\bibinfo{year}{2003}).

\bibitem[{\citenamefont{Shapir et~al.}(1982)\citenamefont{Shapir, Aharony, and
  Harris}}]{SAH82}
\bibinfo{author}{\bibfnamefont{Y.}~\bibnamefont{Shapir}},
  \bibinfo{author}{\bibfnamefont{A.}~\bibnamefont{Aharony}}, \bibnamefont{and}
  \bibinfo{author}{\bibfnamefont{A.~B.} \bibnamefont{Harris}},
  \bibinfo{journal}{Phys. Rev. Lett.} \textbf{\bibinfo{volume}{49}},
  \bibinfo{pages}{486} (\bibinfo{year}{1982}).

\bibitem[{\citenamefont{Koslowski and {von Niessen}}(1990)}]{KN90}
\bibinfo{author}{\bibfnamefont{T.}~\bibnamefont{Koslowski}} \bibnamefont{and}
  \bibinfo{author}{\bibfnamefont{W.}~\bibnamefont{{von Niessen}}},
  \bibinfo{journal}{Phys. Rev. B} \textbf{\bibinfo{volume}{42}},
  \bibinfo{pages}{10342} (\bibinfo{year}{1990}).



\bibitem[{\citenamefont{Kusy et~al.}(1997)\citenamefont{Kusy, Stadler,
  Ha{\l}da{\'s}, and Sikora}}]{KSHS97}
\bibinfo{author}{\bibfnamefont{A.}~\bibnamefont{Kusy}},
  \bibinfo{author}{\bibfnamefont{A.~W.} \bibnamefont{Stadler}},
  \bibinfo{author}{\bibfnamefont{G.}~\bibnamefont{Ha{\l}da{\'s}}},
  \bibnamefont{and} \bibinfo{author}{\bibfnamefont{R.}~\bibnamefont{Sikora}},
  \bibinfo{journal}{Physica A} \textbf{\bibinfo{volume}{241}},
  \bibinfo{pages}{403} (\bibinfo{year}{1997}).

\bibitem[{\citenamefont{Kaneko and Ohtsuki}(1999)}]{KO99}
\bibinfo{author}{\bibfnamefont{A.}~\bibnamefont{Kaneko}} \bibnamefont{and}
  \bibinfo{author}{\bibfnamefont{T.}~\bibnamefont{Ohtsuki}},
  \bibinfo{journal}{J. Phys. Soc. Jpn.} \textbf{\bibinfo{volume}{68}},
  \bibinfo{pages}{1488} (\bibinfo{year}{1999}).

\end{thebibliography}

\end{document}